\documentclass{elsart}
\usepackage{amssymb}
\usepackage{epsfig}

\begin{document}

\begin{frontmatter}

\title{The detection of single electrons by means of a
Micromegas-covered MediPix2 pixel CMOS readout circuit}


\author[A]{M. Campbell},
\author[B]{M. Chefdeville},
\author[B]{P. Colas},
\author[C]{A.P. Colijn},
\author[C,D]{A. Fornaini},
\author[B]{Y. Giomataris},
\author[C]{H. van der Graaf}, 
\author[A]{E.H.M  Heijne},
\author[C]{P. Kluit},
\author[A]{X. Llopart},
\author[D]{J. Schmitz},
\author[C]{J. Timmermans},
\author[C]{J.L. Visschers}
\address[A]{CERN/MediPix Consortium, Geneva, Switzerland}
\address[B]{DAPNIA, CEA Saclay, 91191 Gif sur Yvette Cedex, France}
\address[C]{NIKHEF, Amsterdam, The Netherlands}
\address[D]{Univ. of Twente/MESA, The Netherlands}

\small{Corresponding author: Jan Timmermans, NIKHEF, timmermans@nikhef.nl}

\begin{abstract}

A small drift chamber was read out 
by means of a MediPix2 readout chip as direct anode.
A Micromegas foil was placed 50 $\mu$m above the chip, and electron multiplication
occurred in the gap. With a He/Isobutane 80/20 mixture, gas multiplication factors
up to tens of thousands were achieved, resulting in an efficiency for detecting
single electrons of better than 90\% . We recorded many frames containing 2D
images with tracks from cosmic muons.
Along these tracks, electron clusters were observed, as well as $\delta$-rays.

$Keywords:$ Micromegas, MediPix2, TPC, single electron, pixel, 
pixel segmented anode
\end{abstract}

\end{frontmatter}

\section{Introduction}

Recently \cite{ref:vcipaper} we have demonstrated the possibility
to read out a drift chamber by means of a direct, pixel segmented active anode.
Images of the interaction of $^{55}$Fe quanta with the gas were obtained 
(see Figure~\ref{fig:fe55}) with a
rather low gas amplification factor, since these quanta create some 220 primary
electrons in a small volume of argon gas.
The aim of the research presented in this letter is to prove the feasibility
of the detection of single (drifting) electrons, based on the same pixel
sensor/Micromegas combination.

Our goal is to develop a single-electron sensitive monolithic device $TimePixGrid$
consisting of a CMOS pixel matrix $TimePix$ covered with a
Micromegas \cite{ref:micromegas}. Each pixel will be equipped with a preamp, a
discriminator, a threshold DAC and time stamp circuitry.
Such a sensor would replace the wires (or GEMs, or Micromegas), anode pads,
feedthroughs, readout electronics and cables of TPCs and could generally be
applied in gaseous (drift) chambers. We intend to integrate the Micromegas
grid onto the $TimePix$ chip by means of wafer post-processing
technology ($InGrid$).

In Section~\ref{sec:device} the test chamber including the MediPix2 readout
chip and the Micromegas are described. In Section~\ref{sec:signal}
details on single electron detection and signal development are presented.
Section~\ref{sec:results} describes the data readout and the analysis of
the cosmic ray tracks. It includes a discussion on the observation of 
so-called Moir\'{e} patterns in the detected hit pixel distribution.
The paper ends with conclusions on the present work and an outlook
to our future plans.

\section{The chamber, MediPix2 and Micromegas}
\label{sec:device}
\subsection{The chamber}
\begin{figure}
\begin{center}
\centerline{\psfig{figure=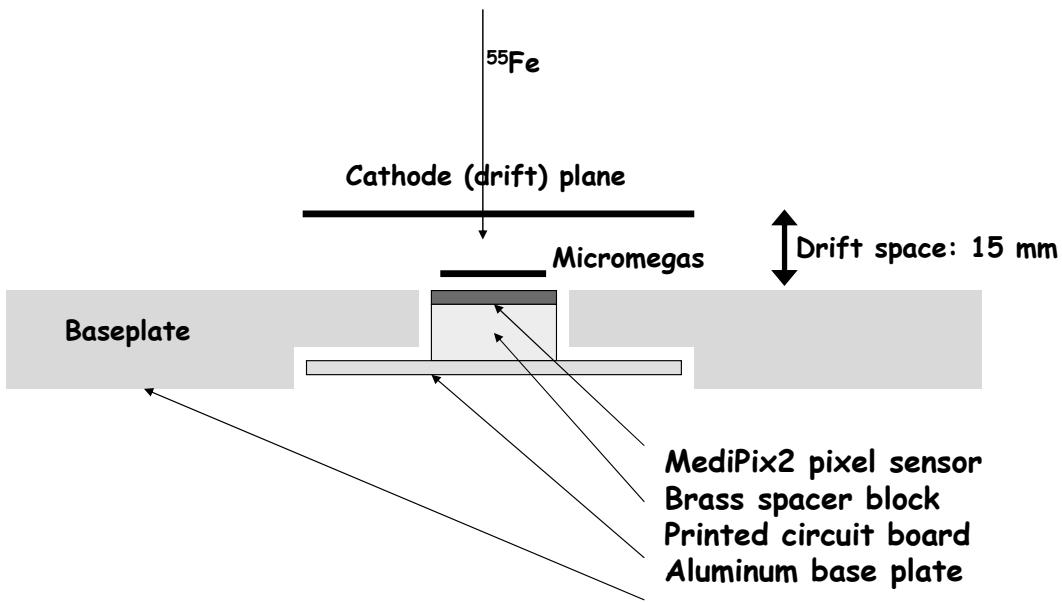,height=14cm,angle=-90}}
\vspace{0.5cm}
\caption{The layout of the chamber with the MediPix2,
the Micromegas and the drift gap.}
\label{detector_layout}
\end{center}
\end{figure}

The test chamber is depicted in Figure~\ref{detector_layout}. 
Above an aluminum base plate, a cathode foil
is fixed, by means of spacers, forming a drift gap of 15 mm height. By means of a cut-out
in the base plate, the MediPix2 chip (mounted on a brass pedestal), is placed flush with
the base plate upper plane. On top of the chip, a Micromegas foil, fixed on a frame, is
held in position by means of two silicon rubber strings 
(see Figure~\ref{detector_photo}).

\subsection{The MediPix2 CMOS pixel sensor}
A MediPix2 chip \cite{ref:MediPix}, \cite{ref:medisoft}, \cite{ref:muros} 
was applied as
experimental readout device. This CMOS chip contains a square matrix
of 256 $\times$ 256 pixels, each with dimensions 55 $\times$ 55 $\mu$m$^2$. Each pixel is
equipped with a low-noise preamp, discriminator, threshold DACs, a 14-bit counter
and communication logic. One edge of the chip has aluminum bonding pads.
The outer dimensions of the chip are 16.12 $\times$ 14.11 mm$^2$.
The MediPix2 chip has been designed for X-ray imaging applications.
For that particular application, an X-ray semiconductor converter (i.e. Si or CdZnTe),
in the form of a corresponding pixel matrix, is mounted onto the MediPix2 chip,
by means of bump-bonding. The assembly of a MediPix2 chip and an X-ray
converter forms a complete X-ray imaging device. For each pixel the number of absorbed
X-ray quanta in a given acquisition time is counted, and the combined pixel content
forms the X-ray image. In our application we use the ``naked" MediPix2 chip,
without an X-ray convertor.

Originally, each pixel of the MediPix2 chip is 
covered for a large part with an insulating passivation
layer; the conductive pad (octagonally shaped, 25 $\mu$m wide) is large enough to
accommodate a bump bond sphere. The electric field in the gap between the MediPix2
and the Micromegas is in the order of 7 kV/mm, 
and discharges were expected when
some 70\% of the anode surface is covered with an insulating material.

\begin{figure}
\begin{center}
\centerline{\psfig{figure=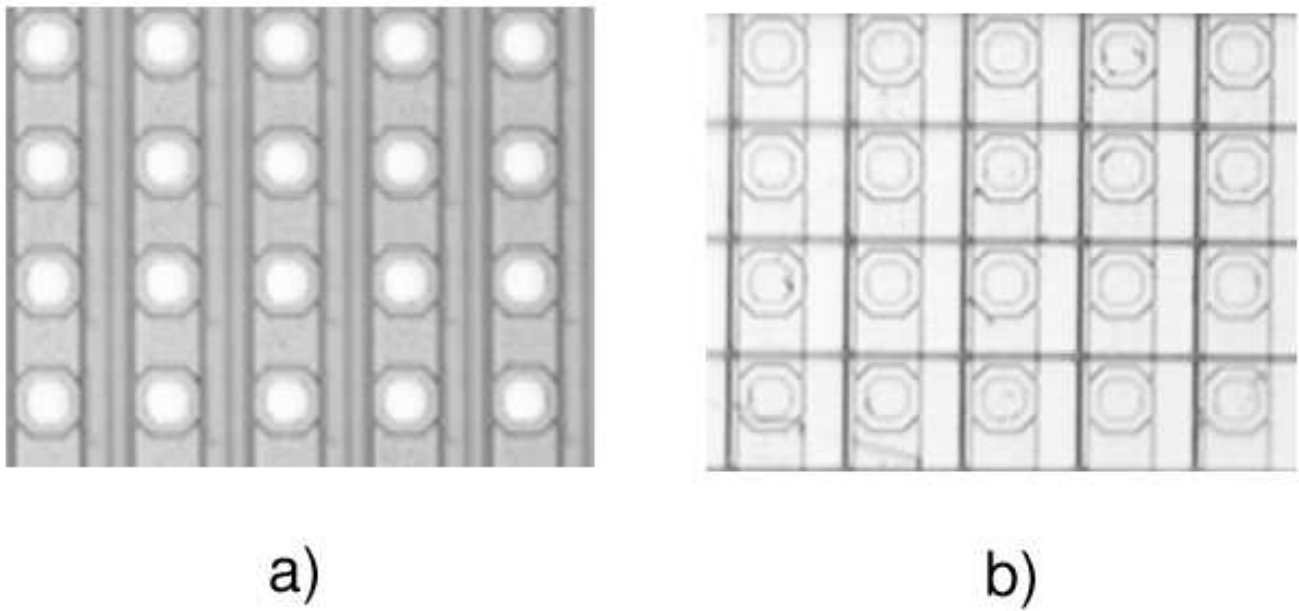,height=8cm}}
\caption{ 
The Medipix2 chip before (a) and after (b) the {\em wafer post processing}. 
The original pad (Al, 25 x 25 $\mu$m) is covered with an 
aluminum pad of 45 x 45 $\mu$m. Note that no Micromegas mesh has 
yet been mounted on the chips shown in this figure.
}
\label{nonmod_mod}
\end{center}
\end{figure}

In order to prevent these discharges, the MediPix2 wafers were post-processed by
MESA+ \cite{ref:mesa}. The post-processing consisted of a deposition of a thin aluminum
layer using lift-off lithography. This allows deposition of metal on the anode matrix
without modification of the bond pads. 
The pixel pads were thus enlarged to reach a metal
coverage of 80\% of the anode plane (see Figure~\ref{nonmod_mod}). 
Electrical tests showed that the
preamplifier functionality was unaffected by this post-processing.
We have applied both the modified and the non-modified versions of the MediPix2 chip.

\subsection{The Micromegas}
The Micromegas is a copper foil, thickness 5 $\mu$m, with holes of
35 $\mu$m diameter in a square pattern with 60 $\mu$m pitch \cite{ref:cernworkshop}.
At the foil side facing the MediPix2 chip, poly-imide pillars
(height 50 $\mu$m, diameter 80 - 140 $\mu$m, pitch (square) 840 $\mu$m) are attached.
The Micromegas, in its frame, was held on the MediPix2 chip by means of two silicon
rubber strings.
When the voltage on the Micromegas was applied (200 - 500 V), the electrostatic force
pulls the mesh towards the chip, and the insulating pillars define the proper gap size.

In order to prevent HV breakdowns, a square kapton foil, with a square hole
of 10.5 $\times$ 10.5 mm$^2$, was placed between the Micromegas and the MediPix2. The chamber
was placed such that the drift direction was vertical. The fiducial drift volume
of 10.5 $\times$ 10.5 $\times$ 15 mm$^3$ is hit by a cosmic ray particle about once
per minute \cite{ref:cosmic}.

\section{Single electron detection; signal development}
\label{sec:signal}
A muon, originating from a cosmic shower, traversing the drift volume, will create
clusters of electrons along its track. The cluster density, and the distribution of
the number of electrons per cluster depends on the gas composition, gas density and
the muon momentum. If argon is the main component of the gas (at atmospheric pressure),
some 3 clusters per mm are created for a minimum ionising muon, and the average
number of electrons per cluster is 2.5 \cite{ref:yellowreport}. Consequently,
on average some 7 primary electrons are created per track length of 1 mm.
The mean distance between two primary electrons, projected onto the
anode plane, is much larger than the pixel pitch, and therefore typically single electrons
will enter a Micromegas hole.
For this reason the single-electron response is essential for the performance of
the pixel-segmented anode readout.
The counting of primary ionisation clusters would allow a precise measurement
of the energy loss dE/dx \cite{ref:digitaltpc}.

After an electron has entered a hole, it will be multiplied, and the number of
electrons grows exponentially towards the anode pads. The centre-of-gravity
of the points of electron-ion separations is positioned at
$D \ln{2}/\ln{M}$ 
away from the anode, where
$D$ is the distance between the Micromegas and the anode and $M$ is the gas
multiplication factor. With $D = 50~\mu$m and $M = 3000$ 
the charge centre-of-gravity is about 4 $\mu$m away from the anode.
The electrons will all arrive within 1 ns at the anode. Most of the ions,
moving much slower, arrive within 30 - 50 ns at the Micromegas,
depending on the gas composition and pressure. 
If a point charge crosses the avalanche gap, then the potentials of both the Micromegas
and the anode change linearly with the distance of the point charge to the anode plane.
The charge on the anode pad below the avalanche is the
sum of the negative electron charge and the positive induced ion charge. 
The fast component has an amplitude
of 10\% of the total charge.
The latter (slow)
component decreases during the drift of the ions towards the Micromegas. 
On adjacent
pads, however, the same ions will induce a positive charge,
which will be at maximum
when the ions are halfway (after some 25 ns) 
between the anode and the Micromegas.
This charge is only a fraction of the avalanche charge, and is back to zero
after the arrival of the ions at the Micromegas.
On these pads we may thus expect a bipolar current signal.


The peaking time constant of the MediPix2 preamp/shaper is 150 ns. This is large
with respect to time constants of the signal development. The preamp/shaper
output is therefore proportional to the avalanche charge,
and the discriminator threshold can be expressed unambiguously in the number of
electrons appearing at the input pad.

Although the design value of the input noise equivalent of the pixel preamps
was 90 electrons, the thresholds were set at 3000 electrons in order to limit
background hits due to (digital) feedback noise, possibly caused by the 
4 mm long bonding wires.

\begin{figure}
\centerline{\psfig{figure=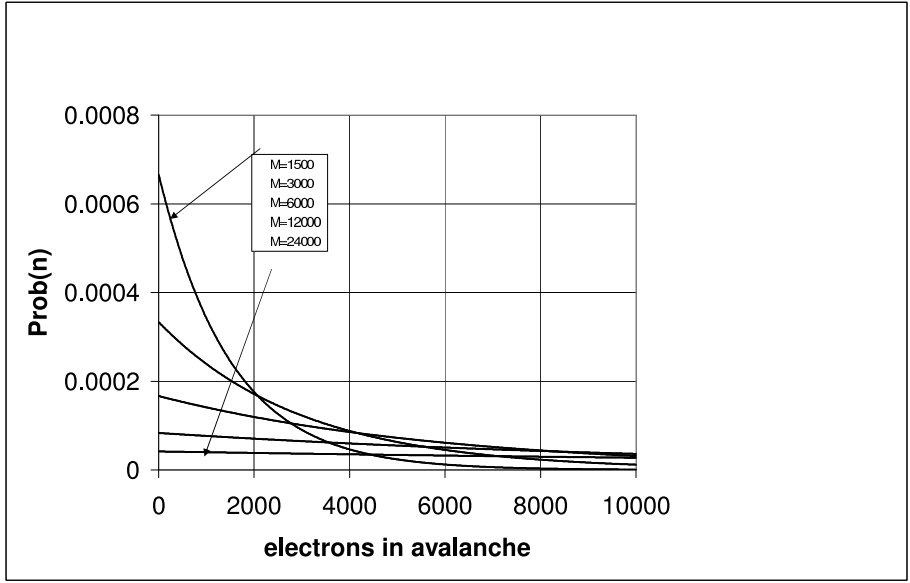,height=12cm,angle=-90}}
\vspace{0.5cm}
\caption{\label{probability}
Probability distribution for the number of electrons in an avalanche
for several values of the gas gain $M$.
}
\end{figure}

The average of the number of electrons in an avalanche, initiated by a single
electron is the gas multiplication factor $M$ \cite{ref:yellowreport}.
The fluctuations in the number of electrons in an avalanche follow an
exponential function \cite{ref:snyder}:

p($n$) = $\frac{1}{M}$  e$^{-\frac{n}{M}}$

where p($n$) equals the probability to have an avalanche with $n$ 
electrons in total.
The avalanche distribution is shown for 
several values of the gain $M$ in Figure~\ref{probability}.
Since the preamp noise is small with respect to the threshold, and since there
is no electron attachment, we apply the simple exponential distribution instead
of approaches like the Polya
distributions \cite{ref:riegler} which include several 2nd order effects.

With a threshold set at $T$ electrons, avalanches smaller than that
are not detected. The efficiency $\epsilon$ to detect single
electrons is then given by:

$\epsilon = e^{-\frac{T}{M}}$

\begin{figure}
\centerline{\psfig{figure=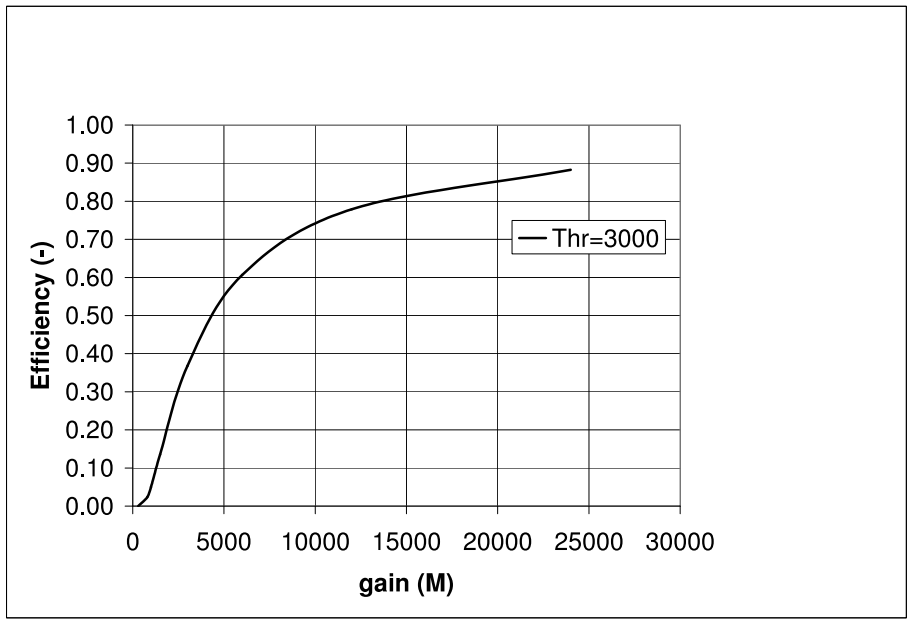,height=12cm,angle=-90}}
\vspace{0.5cm}
\caption{\label{efficiency}
Single electron detection efficiency as a function of the gas gain for a 
threshold set at 3000 $e^-$.
}
\end{figure}

If, for instance, the threshold is set to a value that equals the gain, the efficiency
equals 1/e = 0.37. In Figure~\ref{efficiency} 
the efficiency curve is depicted as a function of the
gain $M$ for a threshold $T = 3000$.
We would like to keep the gas gain as low as possible in order to a) limit the risk
of discharges and ageing and b) limit the ion space charge.
With the present MediPix2, with its lowest threshold
of 3000 electrons, a gain of 10k would correspond to a
single electron efficiency of 0.74.
For this reason we used He mixtures
which allow a high gain, with a small risk of discharges.

Due to discharges, four MediPix2 chips were destroyed 
within 24 hours of operation.
The MediPix2 chip has no protection circuitry at its pixel input pads
other than the source and drain diffusions of the transistors
responsible for leakage current compensation.
We noticed some damage of the pixel
pads, probably due to a high temperature in the discharge region.
For $InGrid$, we intend to eliminate discharge damage by 
a) covering the bottom
of the Micromegas with a (high) resistive layer, 
limiting the participating charge,
b) covering the anode pads with a (high) resistive layer, in combination with
c) a protective network, for each pixel,
connected to the anode pad.

\section{Results}
\label{sec:results}

\subsection{Cosmic ray tracks and data readout; calibration}
The MediPix2 sensor can be externally enabled and stopped, followed by a
readout sequence in which the pixel counter data is transfered to a computer.
We enabled the counters during an exposure time of 15 or 60 s, followed by
recording the image frame in the form of 65k counter contents.
No trigger was applied.

The (positive) charge signals on the Micromegas were read out by means of a low-noise
charge sensitive preamp, with a decay time constant of 1 $\mu$s. Signals from
an $^{55}$Fe source could be recorded, and the preamp was calibrated with charge
signals from a block wave, injected by means of a 10 pF capacitor.
Together with the known
number of primary electrons per $^{55}$Fe quantum, the gas amplification can be measured.

With a He/Isobutane 80/20 mixture, we observed signals 
from $^{55}$Fe events with
$-390$ V on the
Micromegas and $-1000$ V on the drift cathode plane.
This is expected given the large
density of primary electrons in the interaction point \cite{ref:vcipaper}
and the gain
at this voltage of about 1k. We then increased the voltage
on the Micromegas to $-470$ V,
corresponding to a gain of approximately 19k. 
With a threshold setting of 3000 $e^-$, we expect a
single electron efficiency of 0.85, and cosmic rays were observed.

\begin{figure}
\vspace{-1cm}
\centerline{\psfig{figure=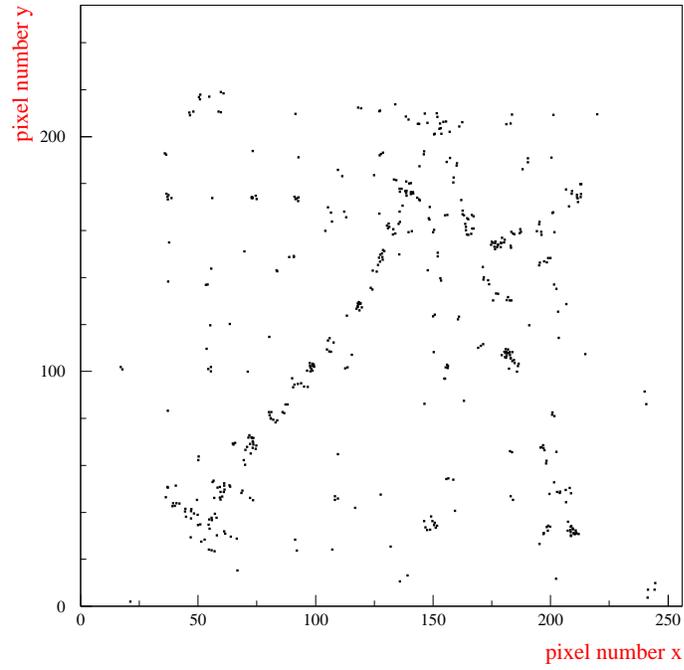,height=10cm}}
\caption{\label{ev1}
Image recorded from the MediPix2/Micromegas prototype TPC showing cosmics
charged particle tracks together with some background. All the hit pixels
during an acquisition time of 15 seconds are shown.
}
\end{figure}

\begin{figure}
\vspace{-1cm}
\centerline{\psfig{figure=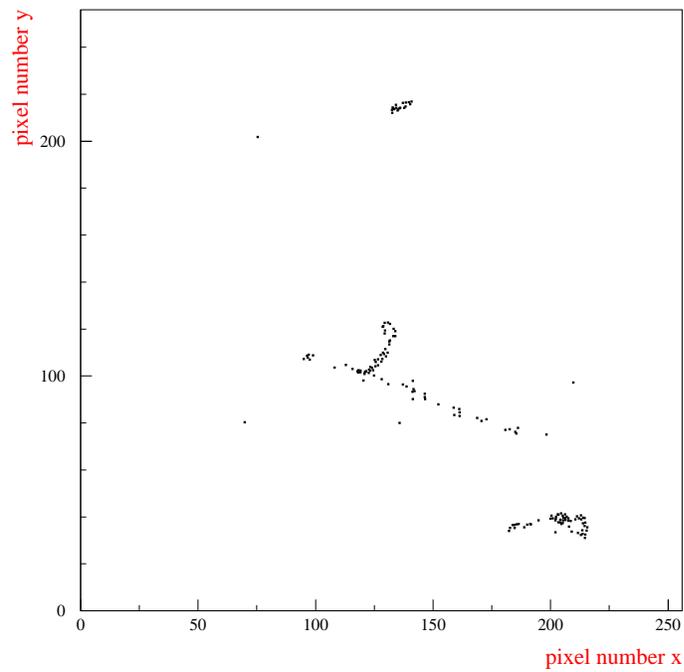,height=10cm}}
\caption{\label{ev2}
Image recorded from the MediPix2/Micromegas prototype TPC showing a cosmic
charged particle track together with a $\delta$-electron.
}
\end{figure}

\begin{figure}
\vspace{-1cm}
\centerline{\psfig{figure=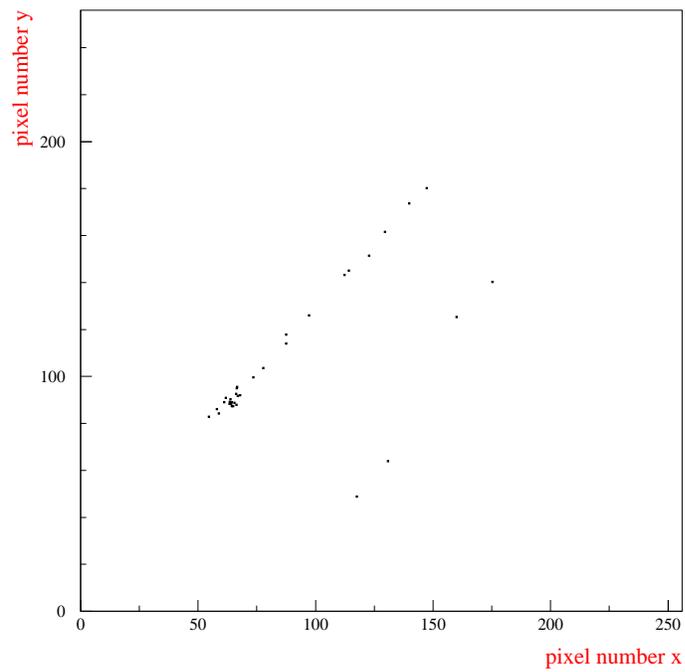,height=10cm}}
\caption{\label{ev3}
Image recorded from the MediPix2/Micromegas prototype TPC showing 
selected cosmics
charged particle tracks. The selections and noise filtering
are described in the text.
}
\end{figure}

Some typical events are shown in Figures~\ref{ev1}-\ref{ev3}.
Figure~\ref{ev1} shows a cosmic event with environmental background.
Figure~\ref{ev2} shows a cosmic muon that knocks out a delta
electron.  Figure~\ref{ev3} shows a selected cosmic muon. 
The selection cuts are described below. In this event
the effect of diffusion can be observed in the spread of the
hits along the track. 

A selection to obtain a sample of clean cosmic events was made.
For the data a map of the noisy pixels was made. The signal from the 
edges and the inefficient
upper left corner of the detector were removed. 
First, a straight line was searched using a Hough transform.
Pixels within a distance of 20 pixels are associated to 
the track.
The following quantities were calculated: 
the number of associated pixels,
the r.m.s. of the distance to the track, 
the track length in the detector plane $L_d$ (in mm).
The full 3D track length $L$ (in mm) is estimated as 
$L=\sqrt{L_d^2+15^2}$. 
The track is split into two equal parts and the
minimum r.m.s.  value of the two parts is calculated.
Clusters are formed by stepping along the track 
and grouping all hits within a distance of 5 pixels.
The number of clusters is counted.  

The following criteria were applied to select an event:
\begin{itemize}
\item{the number of associated pixels larger than 5 and 
the fraction of associated pixels to the track to the total number of pixels 
hit larger than 80\%; } 
\item{ $L_d$ larger than 2.75 mm (i.e. 50 pixels); }
\item{  the r.m.s. less than 4 ; }
\item{  the number of associated pixels per millimeter of 3D track length 
should be less than 4. }
\end{itemize}

\begin{figure}
\centerline{\psfig{figure=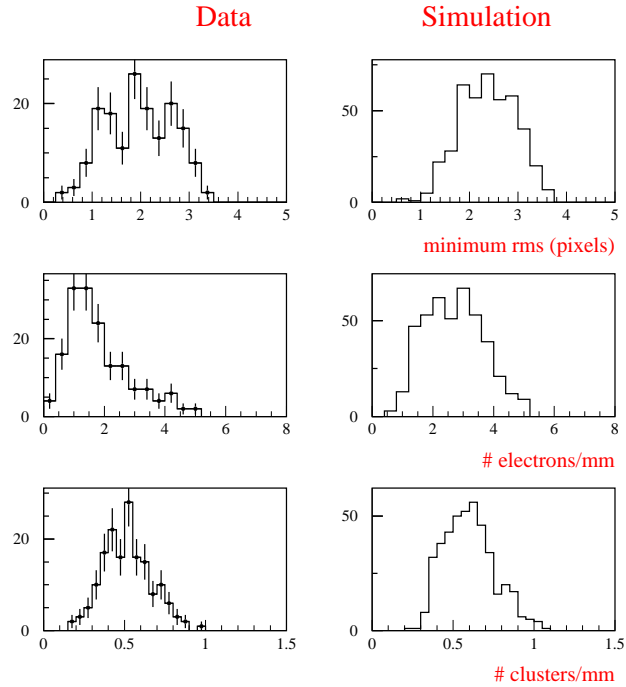,height=9.5cm}}
\caption{\label{ana}
Distributions of some quantities used in the analysis for selected cosmic
data (left) and in the simulation (right). Top: the minimum of the two r.m.s.
values for a track. Centre: the number of reconstructed 
electrons per mm of 3D track length.
Bottom: the number of reconstructed primary clusters per mm of 3D track length.
}
\end{figure}

\noindent
In total 164 events were selected in the data. 
The distributions of some physical quantities are shown in Figure~\ref{ana}.

A simulation programme was written generating cosmic 
minimum ionising particles with an angular distribution 
$\propto \cos^2 \theta$. 
The muon was tracked through the sensitive volume of the detector.
Clusters were generated with an average of 1.4 per mm and 
per cluster 3.16 electrons were generated 
using a Poissonian distribution \cite{ref:magboltz}. 
The electrons were drifted toward the MediPix2 detector with a 
diffusion constant of 220 $\mu$m per $\sqrt{\mathrm {cm}}$. 
Inefficient zones of the MediPix2 detector in the region between 
the pixels and below the pillars were put in. 
The detector is assumed to have an efficiency of 100\% in the efficient zones.
Note that multiple hits on a single pixel are at present not separated.
The same selection cuts were applied to data and simulation.   

The distribution of the minimum r.m.s. is sensitive to the diffusion constant.
Data give an average value of 2.0 pixels (simulation 2.4).
This implies that the diffusion constant is slightly better than 
220 $\mu$m per $\sqrt{\mathrm {cm}}$.
The observed number of pixels hit per mm is 1.83 on average (2.70 simulation). 
The number of clusters per mm is 0.52 (simulation 0.60).
The average 3D track length is 16.5 mm. 
The number of clusters per mm agrees within 15\% with the simulation,
the number of electrons within 35\%.
Note that a 100\% efficiency is assumed for the detector.
Inefficiencies have also more impact on the number of electrons than
on the number of clusters.
If we take into account systematic uncertainties 
on the expected number of clusters and electrons per mm, uncertainties
on the efficiency and operating conditions of the detector, 
we find the agreement reasonable. 
Later experiments will focus on a more precise quantitative understanding
of the detector.


\subsection{Moir\'{e} effects}
Figure~\ref{source1} shows an image, obtained after irradiating 
the chamber with $\beta$'s from a
$^{90}$Sr source. 
The top-left corner of the image is clearly less efficient. This is due to the
non-flatness of the Micromegas in its frame. Apparently, the electrostatic
force could not entirely eliminate the warp in the Micromegas foil.
Here, the pillars are not in contact with
the MediPix2 surface. The gap is wider and the gain is reduced.
The dead regions due to the pillars are clearly visible as well.

\begin{figure}
\vspace{-1cm}
\centerline{\psfig{figure=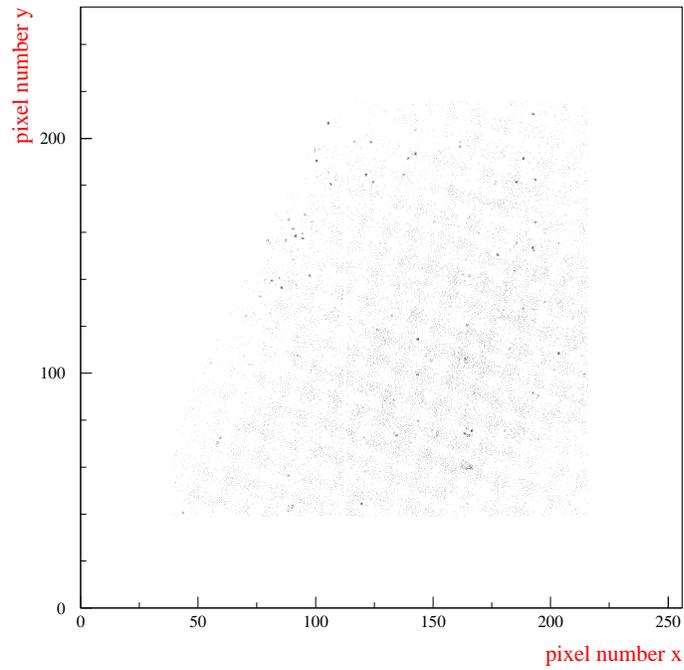,height=10cm}}
\caption{\label{cosmic2}
Superimposed images recorded with a non-modified MediPix2 sensor
during one night of cosmics data taking.
}
\end{figure}

Figure~\ref{source2} shows the image, taken with a non-modified 
MediPix2 sensor,
after irradiating the drift volume with the $^{90}$Sr source. 
Band-shaped regions
with a reduced efficiency are clearly visible.
Note that these bands are present in two perpendicular directions.
The same effect is visible in an image (Figure~\ref{cosmic2})
which is the sum of all cosmic rays obtained during one night of data taking, 
again with a 
non-modified MediPix. 
The corresponding image from a modified MediPix2 is shown in
Figure~\ref{cosmic1}, 
where bands are still present, but much less pronounced.
The bands can well be explained in terms of a Moir\'{e} effect:
the pixel size of the MediPix2 sensor (55 $\times$ 55 $\mu$m$^{2}$) does not
match the pitch of the holes in the Micromegas (60 $\times$ 60 $\mu$m$^{2}$).
Consequently, the hole position with respect to its nearest pixel centre,
shifts when one follows a pixel row or pixel column.
The relative hole position repeats after $\frac{60}{60-55}$ = 12 pixels.

The Moir\'{e} effect can be understood in terms of the pixel signal amplitude
being dependent of the relative position of a Micromegas hole and the pixel
pad. Such a variation could not be explained by the charge sharing
between two or four pixels for avalanches (with a certain width), located
in the region near a pixel edge, because we did not observe the consequent
effect of having significantly more clusters with two, three or four hit pixels
in the same less efficient regions.

Instead, the less efficient regions can be explained by the partly 
insulator-covered anode. If a Micromegas hole is located above the joint
of two pixels, or above the four adjacent corners of four pixels, the drifting
electron will be pulled towards one pad which is relatively far away.
Along this drift path the electric field is less strong, 
and the gain is smaller.
This effect explains the difference in the amplitudes of the Moir\'{e} effect
when using modified or non-modified MediPix2 sensors.


The probability for two-fold clusters was found to be homogeneous, 
and not subject
to the Moir\'{e} effect, 
but the measured value was found to be significantly higher than
the MonteCarlo simulation (30\% and 10\%, respectively).
We explain this by the occurrence of very large but not rare avalanche charges,
following the distribution shown in 
Figure~\ref{probability}. A neighbouring pixel can be hit
due to capacitive crosstalk, in spite of the (positive) induced charge.


\section{Conclusions and Outlook}
We have demonstrated that single electrons can be detected 
with an assembly of a
CMOS pixel chip and a Micromegas foil, with an efficiency larger than
0.9, in a He based gas mixture.
Bubble chamber like images of cosmic ray tracks have been obtained and even
$\delta$-electrons could be observed.
The device allowed to reconstruct the number of primary ionisation
clusters per unit of track length, giving the possibility of a
measurement of the ionisation loss dE/dx.

For the future $TimePixGrid$, 
the grid holes are precisely centered above the pixel
pads, eliminating the non-homogeneity of the efficiency.
The fact that the non-modified MediPix could stand the strong electric field,
together with its strong Moir\'{e} effect makes us confident 
that we can apply
a pixel circuit with small pads, provided that the grid holes are well
centered above the pads. The pad capacity can be kept small, simplifying
the pixel input circuit. 
The inter-pad capacity is then also small, reducing the crosstalk between
neighbouring pixels.
A very small pad may reduce the maximum radiation
dose, due to ageing, and an optimum must be found.

The combination of a pixel sensor and a Micromegas offers an instrument capable
to give a full 2D image of all single electrons in a gaseous volume. 
By replacing
the MediPix2 sensor with a $TimePix$ chip, 
a full 3D image is expected to be within
reach. These circuits will open new possibilities for particle
detection, in terms of position resolution, track separation and energy loss
measurements. As an other example, the polarisation of X-ray quanta can be
measured~\cite{ref:bellazzini}, after its interaction with gas, 
from the direction of the photo-electron,
which is registered accurately with the new device.
Applied with a thin drift space of one mm, the device could be used as a fast
vertex detector in high radiation environments.

\section{Acknowledgements}
We thank the MediPix Collaboration for providing us with several wafers
with MediPix2 chips, for the readout software and hardware. We would like to
thank Arnaud Giganon, Wim Gotink, Joop R\"ovekamp and Tom Aarnink for their
creative and essential contributions to the
realisation of the test detectors.

\newpage

\begin{figure}
\vspace{-2cm}
\vspace{0.5cm}
\caption{\label{fig:fe55}
Image acquired with the Medipix2/Micromegas prototype TPC \cite{ref:vcipaper}. 
}
\end{figure}

\begin{figure}
\begin{center}
\vspace{0.5cm}
\caption{ 
The mounting of the Micromegas onto the MediPix2 sensor.
}
\label{detector_photo}
\end{center}
\end{figure}

\begin{figure}
\vspace{-1cm}
\caption{\label{source1}
Superimposed images recorded from the MediPix2/Micromegas prototype TPC
after irradiation with $\beta$'s from a $^{90}$Sr source. 
}
\end{figure}

\begin{figure}
\vspace{-1cm}
\caption{\label{source2}
Superimposed images recorded with a non-modified MediPix2 sensor
after irradiation with $\beta$'s from a $^{90}$Sr source.
}
\end{figure}

\begin{figure}
\vspace{-1cm}
\caption{\label{cosmic1}
Superimposed images recorded with a modified MediPix2 sensor
during one night of cosmics data taking.
}
\end{figure}











\end{document}